\documentclass{elsarticle}

\usepackage{cite}

\usepackage{amsmath,tikz}
\usetikzlibrary{matrix}
\usepackage{amsfonts}
\usepackage{epstopdf}
\usepackage{color}
\usepackage{algorithm2e}

\usepackage[normalem]{ulem}

\begin{document}
	
	\title{Robust Statistics for Image Deconvolution}
	
	\author[jhucs]{Matthias~A.~Lee}
	\ead{matthiaslee@jhu.edu}
	\author[jhuams]{Tam\'as~Budav\'ari}
	\ead{tamas.budavari@jhu.edu}
	\author[stsci]{Richard~L.~White}
	\ead{rlw@stsci.edu}
    \author[jhuams]{Charles Gulian}
    \ead{cgulian2@jhu.edu}

	\address[jhucs]{Department
		of Computer Science, Johns Hopkins University, Baltimore,
		MD, 21218 USA.}% <-this % stops a space
	\address[jhuams]{Department
		of Applied Math and Statistics, Johns Hopkins University, Baltimore,
		MD, 21218 USA.}% <-this % stops a space
	\address[stsci]{Space Telescope Science Institute, Baltimore,
		MD, 21218 USA.}% <-this % stops a space
	
	\begin{abstract}
		We present a blind multiframe image-deconvolution method based on robust statistics.
		%
		%In this paper we explore the shortcomings of traditional deconvolution methods for removing blur in the presence of artifacts and noise. 
		%
		The usual shortcomings of iterative optimization of the likelihood function are alleviated by minimizing the $M$-scale of the residuals, which achieves more uniform convergence across the image.
		We focus on the deconvolution of astronomical images, which are among the most challenging due to their huge dynamic ranges and the frequent presence of large noise-dominated regions in the images. We show that high-quality image reconstruction is possible even in super-resolution and without the use of traditional regularization terms.
		%
		%This is a particularly difficult problem due to the inherently high dynamic range and the presence of large noise dominated areas.
        %
        Using a robust $\rho$-function is straightforward to implement in a streaming setting and, hence our method is applicable to the large volumes of astronomy images. %, approach guided by Robust statistics. 
		% Charlie: I'm not sure if this sentence is coherent but perhaps I need more background-- what about a robust rho-function lends itself to a streaming setting? Perhaps separate these two thoughts: 1) We use a robust rho-function to improve convergence of faint sources, and 2) Because our algorithm can be implemented in a streaming setting, it is useful in the context of large astronomical surveys such as LSST.
		%
		The power of our method is demonstrated on observations from the Sloan Digital Sky Survey (Stripe 82) and we briefly discuss the feasibility of a pipeline based on Graphical Processing Units for the next generation of telescope surveys.
	
	\end{abstract}
	
	%\begin{IEEEkeywords}
	%	Robust Statistics, Deconvolution, Streaming, Super Resolution, Astronomy
	%\end{IEEEkeywords}
	
	%\IEEEpeerreviewmaketitle
	
	\maketitle
	
	\section{Introduction}
	
	In the new era of astronomy surveys, dedicated telescopes observe the sky every night to strategically map the celestial sources. The next-generation surveys are capable of such high speed that repeated observations become possible, opening a new window for research to systematically study changes over time. The key requirement for time-domain astronomy is the development of sophisticated algorithms that can maximize the information we gain from the data. The image processing approaches we present in this paper are motivated by these astronomical challenges but are not specific to such exposures and are expected to work for long-range photography regardless of the content of the images.
	
	Algorithmically deblurring single telescope images has had a long history. At first during the 1970s, when \citet{richardson1972bayesian} and \citet{lucy1974iterative} independently introduced a Poisson-based iterative deconvolution method, generally known as the Richardson-Lucy, and then during the early days of the Hubble Space Telescope when researchers rushed to remove the blur induced by the misaligned optics \citet{burrows1991imaging,cunningham1993image,krist1993deconvolution,lucy1992co,nunez1993general}. Most of these techniques rely on a variation of an iterative deconvolution using a \emph{known} point spread function (PSF). The main difficulty with such methods is preventing numerical degeneracies caused by the presence of noise and poorly constrained parameters \citep{homrighausen2011image}. Even more challenging is when the PSF is unknown. Some techniques, such as the iterative non-negative matrix factorization (NMF) approach described in \citet{fish1995blind}, have shown limited success simultaneously solving for the PSF and the latent image. The aforementioned approaches are limited by the total amount of information contained in the input image and the PSF (if available).
	Ground-based imaging, astronomy or long-range photography a like, suffers from varying distortions introduced by turbulence in Earth's atmosphere. Multiple exposures, however, provide an opportunity to break the degeneracy of the varying PSFs and the constant background image.
	Blind multiframe deconvolution uses multiple images to simultaneously solve for the latent image and the individual PSFs. This blind deconvolution method is based on Fish's approach of iteratively solving for both the PSF and the latent image, the difference being that new observations are regularly introduced in order to add more information into the deconvolution, this is the basis of the approach described in \citet{harmeling2009online}.
	
	The current state-of-the-art techniques in astronomy for extracting and combining repeated observations are relatively simple methods when compared to other areas of image processing. This is in part due to the incredibly high dynamic range of the images, as well as the occurrences of large noise dominated area which tend to handicap methods from other areas. The two most common techniques used are \emph{lucky imaging} \citep{tubbs2003lucky} and the \emph{coadding} of observed images \citep{lucy1992co,annis2011sdss}. Both of these use linear combinations of the input images, which enables proper noise propagation, although the full co-variance matrix is rarely produced.

	\emph{Lucky imaging} is the process of only choosing the observations with the very best PSF and throwing away the rest, often over $90\%$ of the original data is discarded. This yields sharper images, but with a low signal-to-noise ratio. 
	A \emph{Coadd} image is the combination of multiple images, produced by generating a pixel-by-pixel mean or median image, therefore suppressing noise and outliers. 
	However, considering that the input images have different PSFs, one needs to first convolve them to match the worst acceptable blur before combining the pixel values. This results in a high signal-to-noise ratio but also in an overall blurrier solution than the majority of input images.
	In practice, usually a combination of both of these methods are applied; only the best images are chosen and then co-added, meaning that a large amount of potentially useful data is discarded in the process.

	In a perfect world we would like to clean up every single observation without throwing away any useful data, and restore each of them to pristine condition with infinite resolution, but this is impossible. It is, however, possible to extract and combine information across all observed frames and produce a few sharper images of higher resolution, exposing sources and features previously hidden in blur and noise. True reconstruction is computationally complex and therefore slow and laborious. In order to keep up with the growing volume of data, we need fast, statistically sound tools to explore and deblur these images in near real time.
	
	\begin{figure}
		\centering
		%\leavevmode
        \makeatletter
        \let\@currsize\normalsize
		\includegraphics[width={0.6\linewidth}]{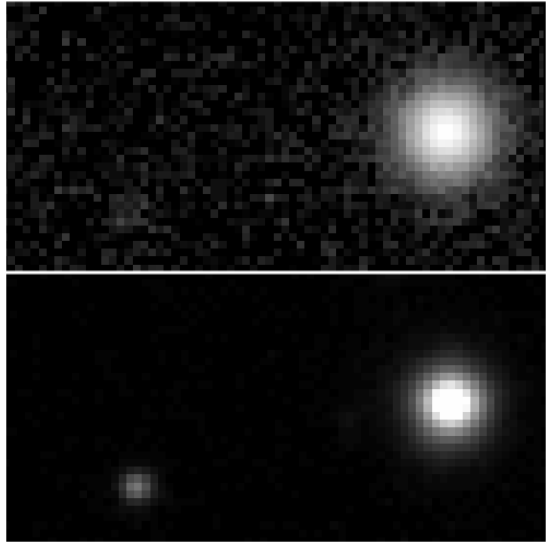}
		\caption{Shown above is an example of Sloan's Stripe 82 observations, displaying the problematic features found in these images. \emph{Top} shows an observation with low signal-to-noise and a large PSF, \emph{Bottom} same section of the SDSS Coadd \citet{annis2011sdss}, showing improvement in signal-to-noise and definition of sources over the plain observation.}
		\label{coadd}
	\end{figure}
	
	For the purpose of this paper we demonstrate the application of our method on the Sloan Digital Sky Survey (SDSS) \citep{york2000sloan}, 
	which over a span of 7 years imaged a large part of the southern hemisphere  with roughly 80 fold coverage. This area known as \emph{Stripe 82} \citep{abazajian2009seventh} is an ideal testbed for demonstrating the power of our new methodology. 
	The overall quality of Stripe 82 images varies widely, see Fig.~(\ref{coadd}). Some images are blurry, some are faint and noisy, but most are a combination of those. \citet{annis2011sdss} produced a state-of-the-art coadd of this region, which we use as a reference in our experiments. Due to the extremely high dynamic-range we encourage the reader to view the included images on a computer screen, as print materials have difficulty reproducing the full range of contrast. 
	
	In this paper, we present a novel approach for extracting information from atmospherically distorted repeated observations in order to produce a deblurred, low-noise, higher resolution image. In section \ref{sec:approach}, we discuss previous work as well as the general approach. In section \ref{sec:improvements} we introduce our robust improvements. In section \ref{sec:application} we discuss the application, results and performance of our method and finally, in \ref{sec:summary} with conclude with future work and a summary of our achievements.
	
	\section{Deconvolution as likelihood optimization}\label{sec:approach}
	
	Before we explore removing the blur, we must formally understand how the blur is induced into our observations. 
	For the mathematical description we consider one-dimensional images represented by (column) vectors but all derivations and equations are similar and apply in higher dimensions as well. In particular here we will focus on 2-D images.
	
	Our model is relatively simple: Each observation $y_t$ at time $t$, is modeled as the convolution of the underlying \emph{true} image, $x$ and the PSF $f_t$, on top of which there is measurement noise $\epsilon_t$,
	\begin{equation}
	y_t = f_t \ast x + \epsilon_t \ .
	\label{basiceq}
	\end{equation}
	Our goal is to fit this model to all observations in time $\{y_t\}$. Going forward we will make the differentiation between the underlying \emph{true} image, $x$, and its estimate, $\tilde{x}$, likewise we will distinguish between the observation, $y_t$, and our reconstruction, $\tilde{y}_t$. Considering that a convolution is a linear operation, we can represent the 1D convolution with $f$ by a matrix $F$, such that \mbox{$Fx\!=\!f\!\ast{}\!x$}, this also holds true for the equivalent 2D convolution. Hereafter we use this convention to representing matrices with capital letters and vectors with lower case symbols.

	%\subsection{Deconvolution as likelihood optimization} \label{sec:deconv}
	%
	The literature features a variety of methods for deblurring with a known point-spread function (PSF). Especially common are methods which maximize the Poisson likelihood, such as the Richardson-Lucy \citep{lucy1974iterative,richardson1972bayesian,starck2002deconvolution} deconvolution or the more noise resilient damped-Richardson-Lucy \citep{white1994image}. There are also blind methods, e.g.,  \citet{fish1995blind}, which do not require a \emph{known} PSF, but instead solve for it as part of the iteration. This is a crucial feature as in most applications, PSFs are not inherently known.
	
	While the noise in CCD observations follows a Poisson distribution, cf.~the number of electrons in CCDs proportional to the number of photons in each pixel, its applicability is often limited because creating a complete model for an image is not practical due to many known and unknown contributions (gradients from moon, thin clouds, etc.)
	The image processing pipelines correct and calibrate the input images. Estimates of the sky background are subtracted, and while noise estimates for each pixel remain accessible, the transformed pixel values will have different noise properties. 
Fortunately the background counts in typical images are high enough that a Gaussian likelihood is a good approximation to the likelihood function.
Our approach originates in Bayesian statistics but full inference is computationally too expensive and hence we resort to maximum likelihood estimation (MLE).
\iffalse
%
The Gaussian error in the pixels means that MLE maps onto $L_2$ optimization,
	\begin{equation}
	\tilde{x} =\textrm{arg}\min_x\sum_t \sum_p \left[y_{t,p} -  F x\right]^2
	\ .
	\end{equation}
	\begin{equation}
	\tilde{x} =\textrm{arg}\min_x\sum_t \sum_p w_{t,p} \left[y_{t,p} - \tilde{y}_{t,p}(x) \right]^2
	\ .
	\end{equation}
\fi
The multi-frame blind deconvolution method described by \citet{harmeling2009online,harmeling2010multiframe} uses a quadratic cost function, which corresponds to the Gaussian limit.
	The resulting formula is similar to that of the Richardson-Lucy. 
	We can solve for the model image, $\tilde{x}$, which minimizes the residuals in all pixels,
    \begin{equation}
	\tilde{x} =\textrm{arg}\min_x\sum_{\textrm{pixels}} \left[y -  F x\right]^2
	\ .
	\label{minimization}
	\end{equation}

	In order to solve for the $\tilde{x}$-image update we use an adaptation of Harmeling's multiplicative update formula
	\begin{equation}
	\tilde{x}_{t+1} = \tilde{x}_t \odot u_t 
	\label{updateImageModel}
	\end{equation}
	with 
	\begin{equation}
	u_t = \frac{F^T W y_t}{F^T W F\tilde{x}_t}
	\label{updateFormula}
	\end{equation}
	where $\odot$ symbol and the fraction indicate element wise multiplication and division. We consider the fraction to be the update image, $u_t$, derived from the observation $y_t$, which defines how adjustments are introduced into the model image. 
	This update image is then applied to our model, as shown in \ref{updateImageModel}. 
	
	There are multiple ways to solve for the PSF, \citet{harmeling2009online} proposes using the LBFGS-B Algorithm \citep{byrd1995limited}, Homrighausen proposes a ``Fourier deconvolution'' method \citep{homrighausen2011image} and \citet{fish1995blind} proposes simply using the same update formula for the estimation of the PSF as for updating the model image. The latter is possible since the convolution, $f\!\ast{}\!x$, is commutative, meaning that if we can use an update formula to solve for the model, we must also be able to rewrite it for the PSF. To do so, we simply interchange the occurrences of the PSF and the model in our formula. Essentially we alternate holding the $f$ and the $\tilde{x}$ constant while estimating and updating the other. For the purposes of this paper we assume the PSF to be constant across our images, but these methods could be extended to use multiple separate PSF or use methods such as Lauer's approach to spatially-variant PSFs \citep{lauer2002deconvolution}.
	
	To regulate our solutions, we constrain both $\tilde{x}$ and $f$ to be non-negative and initialize them to non-zero values. While it is possible to initialize $\tilde{x}$ to a constant value, we found initializing with the average of a few observations speeds up convergence and yields better final results with fewer processed observations.
	
	It is important to prevent erroneous zero values from cropping up in our model and PSF, as once an element becomes truly zero, no future multiplicative update will be able to change it. This is especially tricky as our intended background, for our model image, is as close to zero as possible without exactly reaching it. Floating point underflow can be part of this problem and therefore needs to be dealt with appropriately.
	
	In practice a number of problems occur due to random noise in the data and numerical degeneracies. In the following sections we describe our novel methods for overcoming these limitations.

	\section{Improving stability and convergence}\label{sec:improvements}
	\subsection{Robust statistics}\label{sec:robust}
	Arriving at a good estimate for the PSF image is crucial for successful deconvolution. With the unmodified update formula, the cost function that is being minimized is an $L_2$ norm, therefore featuring a squared term. Extreme outliers in the residual, especially early on in the estimation before the PSF has been well formed, can overtake the residual, forcing convergence in those areas before the rest of the image. This often leads to PSFs more resembling of a \mbox{Dirac-$\delta$} than a realistic PSF.
	
\begin{figure}
		\centering
		%\leavevmode
		\includegraphics[width={0.6\linewidth}]{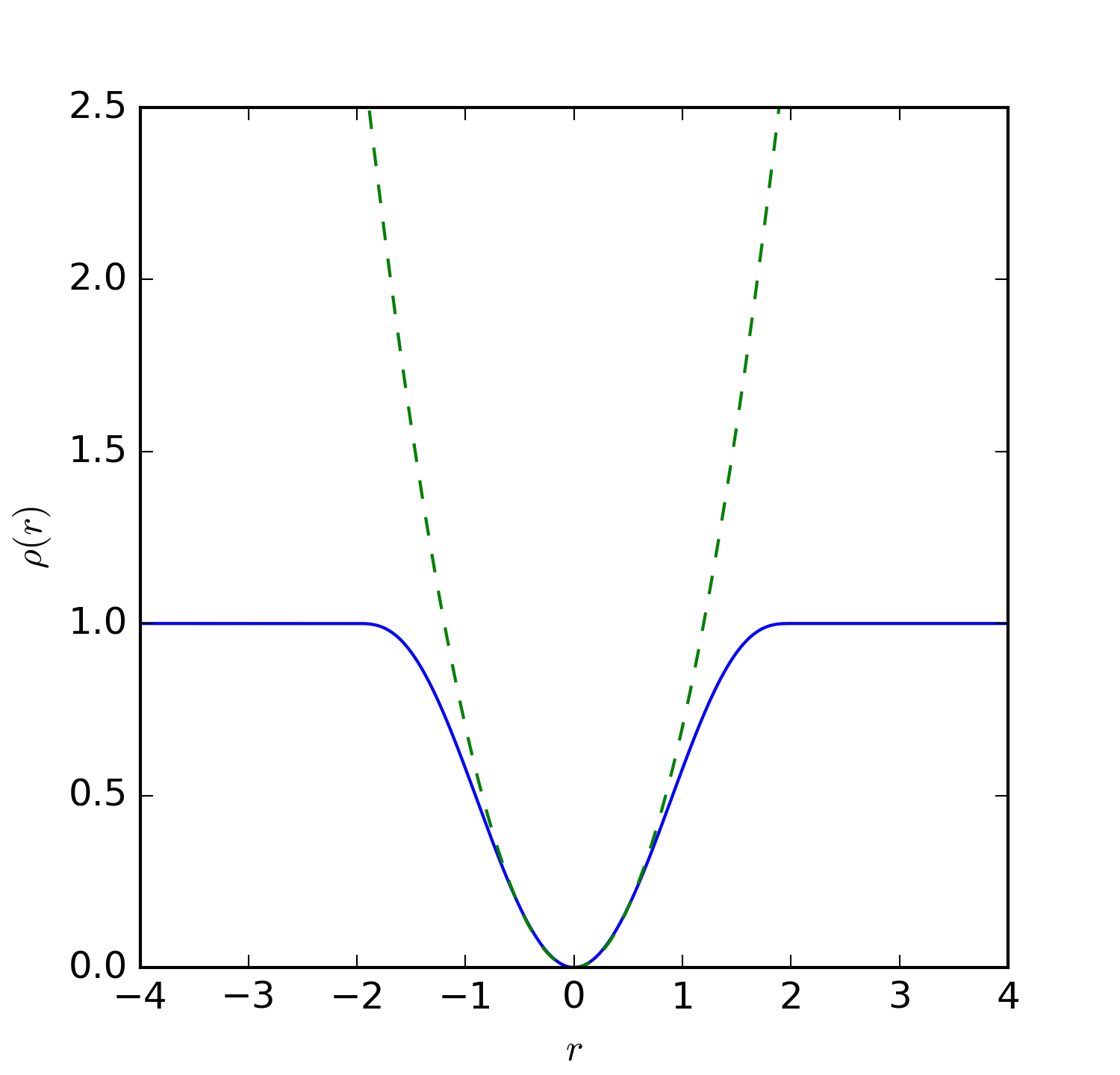}
		\caption{\emph{$\rho$-function} associated with the \emph{bisquare} family of functions. Note the dotted line indicates the contribution of a purely quadratic function.}
		\label{fig:pfunc}
	\end{figure}

	To curb the impact of those very large values, we borrow elements of robust statistics and modify our quadratic cost function to be an $M$-estimator with a robust $\rho$-function \citep{maronna2006robust}. The solution is still iterative in nature and hence lends itself well to our method. We adjust our cost function with the $\rho$-function, 
	\begin{equation}
	\tilde{x}=\textrm{arg}\min_x\sum_{\textrm{pixels}} \mathbb{\rho}\left(y - Fx\right) 
	\ .
	\label{minimizationRho}
	\end{equation}
	In practice the solution to this general problem of robust statistics is obtained by minimizing a weighted $L_2$ problem
	\begin{equation}
	\tilde{x}=\textrm{arg}\min_x\sum_{\textrm{pixels}} w\,\left[y - Fx\right]^2 
	\ ,
	\label{minimizationWeight}
	\end{equation}
	where the weight of each pixel $w$ is computed for its residual based on the current solution of plugging $\tilde{x}$ into \mbox{$\mathbb{W}(r)\!=\!\rho'(r)/r$ function \citep{maronna2006robust},}
	\begin{equation}
	w = \mathbb{W}(y-F\tilde{x})
	\end{equation}
	In particular we apply the \emph{bisquare} function, also known as the \emph{Tukey bi-weight}, \mbox{$\mathbb{W}(r)\!=\!\min\big\{3\!-\!3r^2\!+\!r^4,\,1/r^2\big\}$}, which corresponds to a $\rho$ -function that is quadratic for low values but approaches a constant for large arguments, essentially limiting the contribution of the largest residuals in the cost function.
	Figure~\ref{fig:pfunc} illustrates the difference between the terms in our robust cost function.
	\color{black}
	For more details we refer the interested reader to the discussion of the \emph{bisquare scale} in \citep{maronna2006robust}. 

	The threshold at which the $\rho$-function deviates from being quadratic, is tuned by scaling the residuals, which therefore controls where the dampening starts. In order to relate this tuning parameter to the quality of the image, it is natural to define this scaling in units of $\sigma$.
	This controlled dampening of the residuals greatly increases the quality of the estimated PSFs and therefore also the resulting image. More detailed results are presented in section \ref{sec:applicationsdss}. 
	
	\subsection{Convergence}\label{sec:clipping}
	
	The addition of the weighting as described in the previous section helps constrain the PSF estimation and therefore the areas around objects. On the other hand, the noise-dominated areas \emph{between} objects remain under-constrained, resulting in the occurrence of background artifacts.
	
	The cause of these artifacts had been a long standing problem for this method, until we noticed that as the deconvolution approaches convergence, the useful and reasonable updates to our model image become smaller, more sporadic and clustered around sources. Conversely, the updates in the regions between sources, where values are already low and close to zero, can become extreme due to noise present in the observed image. For example, if a background pixel in our model is near zero, but noise in the observed image wants to push the value higher, update factors of 1000x or more are not uncommon. In most cases the resulting pixel value will still be very small and will fluctuate around zero, cancel each other out. Unfortunately in some scenarios these persist as artifacts, ie. bright speckles across the noise dominated areas of the image. 
	
	To prevent these updates from affecting the model image we introduce an update clipping function, 
	\begin{equation}
	u_t' = \max \big\{1/d,\,\min(d,\,u_{t})\big\} % \quad \textrm{where} \ d>1
	\label{updateFunc}
	\end{equation}
	where $d>1$, which limits the maximum impact an update is allowed to have on any single pixel. The closer the parameter $d$ is to 1, the higher the impact and therefore the more conservative the updates will be. When $d$ is large, the clipping has virtually no impact. This approach vastly cuts down the number of background artifacts. 
	
	Limiting updates in such a way has no real drawback on the dynamic range in practice. For \mbox{$d\!=\!2$} the contrast becomes $2^{40}$ with only 20 iterations.
	
	\begin{figure}
		\centering
		%\leavevmode
		;		\includegraphics[angle=-90,width={0.85\linewidth}]{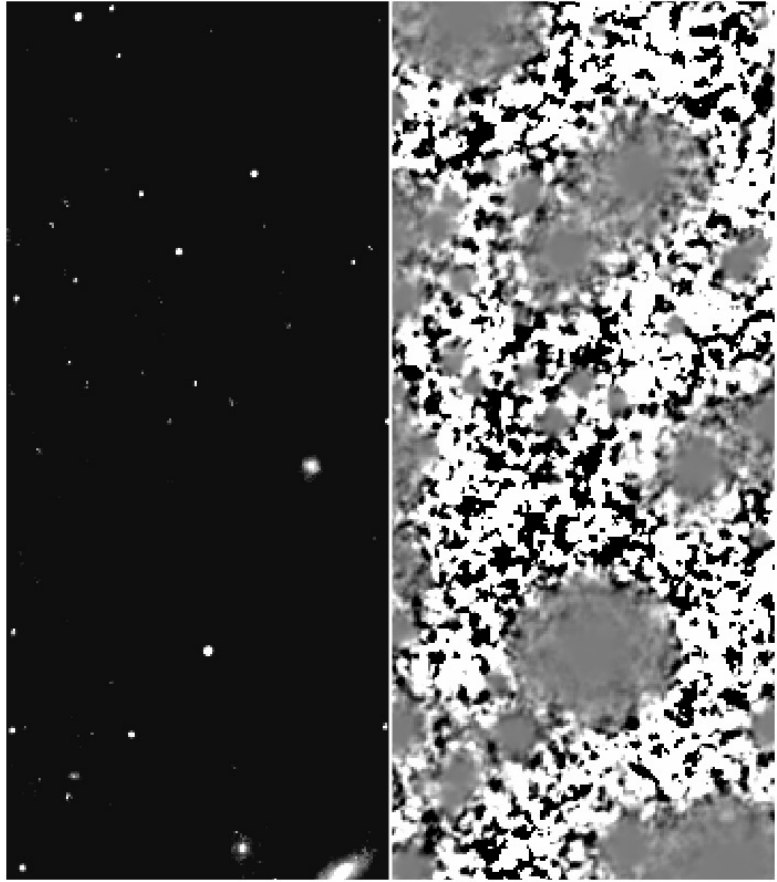}
		\caption{A typical update (\emph{right}) to our image model (\emph{left}) contains the most reasonable updates clustered around objects. The values between objects, the background, are already near zero, so any noise or non-uniform background subtraction can produce extreme and unreasonable update values. Note: the coloring of the update image is such that the areas that fall below the bottom clipping are colored black and the areas that are above our clipping range are colored white. The gray areas are values which we consider to be reasonable.}
		\label{fig:updateImage}
	\end{figure}
	
\begin{figure}
		\centering
		%\leavevmode
		\includegraphics[width={0.85\linewidth}]{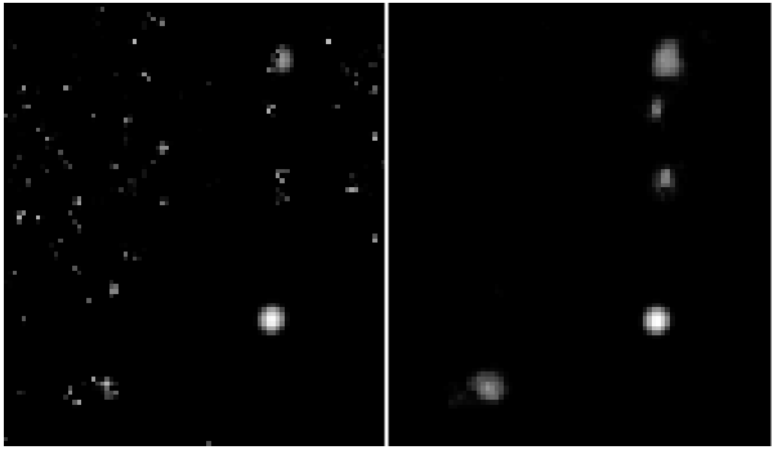}
		\caption{Updates to our model image fluctuate most in areas between sources, without dampening these updates, speckles (\emph{left}) get introduced into the noise-dominated background and faint sources can get disrupted by extreme erroneous updates. By limiting the absolute magnitude of these updates we get much more coherent result (\emph{right}).}
		\label{fig:updateClipping}
	\end{figure}

	In fig.~\ref{fig:updateImage} we show an update image clipped with $d=2.0$ (\emph{right}) and the corresponding current model estimate (\emph{left}). Observe how the gray areas, where the update is between 0.5 and 2.0, fall directly around the regions where objects are located in the current estimate. The areas that either appear pure white or black are where the update was originally above 2.0 or below 0.5 and therefore have been clipped respectively, meaning we only allow each pixel to be either doubled or halved with any update. The resulting effect of this clipping can be seen in fig.~\ref{fig:updateClipping}, the left image shows some typical artifacts, a speckled background in noise dominated areas, as well as the adverse effects of extreme erroneous updates to faint objects. On the right, is the result of an identical deconvolution with the Update Clipping enabled. 
	
	\section{Application}\label{sec:application}
	In the following sections we describe the more traditional components of our method, as well as the overall algorithm. We present the results of our method when applied to real data and also quickly discuss our technical implementation and it's performance. 
	
	\subsection{Pixel censoring}\label{sec:masking}
	Our linear model cannot represent omnipresent artifacts such as saturated pixels, cosmic rays, etc.
	Masking these pixels allows us to process real images containing these artifacts.
	We introduce binary \emph{mask} images that zero out the known-bad areas and leave the remainder untouched.
	Each incoming observation $y_t$, has its own mask associated with it, which for our tests had been pregenerated using masking data available from SDSS.  
	Formally the mask application can be written as a matrix multiplication with a diagonal matrix $W$, which enters the update formula as follows, 
	\begin{equation}
	u_t = \frac{F^T Wy_t}{F^TWF\tilde{x}_t}
	\ .
	\label{eq:masking}
	\end{equation}
	
	Relatedly flux from objects located just beyond the edge of the image as well as the abrupt edge of zero padding used with FFTs can cause artifacts along the mask's edges, which over multiple iterations can creep into the PSF and begin corrupting the model. To prevent these artifacts, we taper the masks towards zero around the edges and any masked out object, therefore smoothing out these artificial hard edges.
	
	\subsection{Super Resolution}\label{sec:superres}
	Access to multiple images of the same objects, also gives us the ability to increase the spatial resolution beyond that of the source images.
	Based on the nature of our images, we can expect each to have a slightly different relative sub-pixel shift, meaning we can extract and combine that information into a higher resolution image model. 
	To do so, we introduce a \emph{downsampling} operator $B$ that lowers the resolution of an image. If the model image $\tilde{x}$ has a factor of 2 higher resolution than the observations $y$, then $B$ would halve the resolution of $\tilde{x}$ in order to make it comparable to $y$ for calculation of the residual. We can model this relation as
	\begin{equation}
	y \approx B F \tilde{x}
	\end{equation}
	Similarly an \emph{upsampling} operator can also be introduced, which is formally the transpose, $B^T$, such that $BB^T$=$I$. These operators are easy to visualize when $B$ and $B^T$ act upon a vector. For a simple vector of length 3, the appropriate operators would be as follows,
	
		\begin{equation}	
		\ \ B=\frac{1}{\sqrt{2}}\begin{tikzpicture}[baseline=(current bounding box.center)]
		\matrix (m) [matrix of math nodes,nodes in empty cells,right delimiter={]},left delimiter={[} ]{
			1 &1 & & & & \\
			& & 1 &1 & & \\
			 & & & & 1 & 1 \\
		};
		\end{tikzpicture} \ .
		\end{equation}
	
	In practice, however, there is no need to create these matrices, we simply implement an operation which doubles each pixel (in the 2x super resolution case) along both dimensions and therefore grows the image. The inverse operation shrinks the image by a binning along each dimension, gathering up the flux that was spread between pixels in the upsampling operation. Each upsampling and downsampling operation is normalized such that the total sum of the fluxes stays constant.

	The addition of super resolution significantly increases the quality of our result; many small sources which in standard resolution resolve to small jagged/aliased groups of pixels, become uniform in shape and better defined. Higher resolution also allows us to more precisely measure sizes and distances between objects.
	
	\subsection{Assembling the pieces}
	The overall method proceeds as shown in Algorithm \ref{algo1}. For every observation, $y_t$, we initialize a new PSF and load any associated masks. We then iteratively estimate a new PSF, while holding our image model constant. During this iteration, we apply the aforementioned Robust Statistics weighting, repeatedly updating the our PSF estimate until we've reached convergence. We use two criterions to measure the PSF convergence, an absolute, which quantifies the per-pixel change by calculating a root-mean-square of the differences between the current and previous estimate, and a relative, which ensures we stop once the maximum relative per-pixel change drops below $0.01\%$. We stop the convergence once either of these criterions are violated. Once we have an acceptable solution, we compute the update for the image model, while in turn holding the PSF constant. The image model is only updated once per iteration, again we apply our Robust Statistics weighting and our Update Clipping to prevent extreme updates from having adverse effects on our model image. Since the resolved PSFs can be wildly different between observations, the only piece of information carried from one observation to the next is the image model.

	\begin{algorithm}
		\SetKwInOut{Input}{Input}\SetKwInOut{Output}{Output}
		\Input{Repeated exposures, $\{y_t\}$}
		
		Initialize latent image $\tilde{x}$ and the exposure masks\;
		
		\For{each new exposure $y_t$}
		{
			Initialize PSF $\tilde{f}_t$\
			%			Load mask for current $y$ and taper edges\;
			
			%			\While{residual \textgreater \ convergence threshold}
			\While{not converged}
			{
				Improve $\tilde{f}_t$ using robust statistics\;
				%				Apply Robust Statistics Weights to update\;
				%				Apply update to PSF\;
			}
			
			Solve for update $u_t$ with robust weights\;
			%Update image using estimated PSF\;
			%			Apply Robust Statistics weights to update\;
			Apply clipping for stability of sky pixels\;
			Update estimated latent image $\tilde{x}$\;
		}
		\Output{Latent image and all PSFs}
		\ 
		\caption{Robust blind deconvolution}
		\label{algo1}
		\color{black}
	\end{algorithm}

	\subsection{Application to SDSS}\label{sec:applicationsdss}
	As mentioned previously, we apply our method to repeated observations of SDSS's Stripe 82. We further restrict our study to the region where Stripe 82 overlaps with the Canada-France-Hawaii Telescope Legacy Survey (CFHTLS), featuring a much larger telescope therefore yielding a much deeper image. 
	This allows us to make comparisons not only to the SDSS coadd but also the CFHTLS coadd by \citet{hudelot2012t0007}. CFHTLS has nearly double the angular resolution of SDSS (SDSS pixel size 1px=0.396", CFHT pixel size 1px=0.187"), making it an ideal candidate for comparison of our super resolution results. CFHTLS enables us to verify our results acting as a sort of \emph{ground truth}. The particular area we chose yields 68 usable observations. All comparisons figures in this paper are shown on a log scale to highlight robust performance even close to the sky background and around faint objects. For fairness of comparison we scale all our images within each figure to the same total flux, contrast and bias.
	
	%\subsection{Beyond Coadds}
	In combination, the previously described methods allow us to produce superior results far exceeding the SDSS coadd in quality and even more impressively the results of the CFHTLS coadd. In this section we will discuss our results and the parameters used to achieve them. 
	Good masking is the basis for all of our results, we generate masks as described in section \ref{sec:masking}. The edges are then softened using a $15$ pixel Gaussian blur preventing edge artifacts.
	The next ingredient is our robust statistics weighting, section \ref{sec:robust}. We experimented with a large variety of different values for the tuning parameter and found $T=6$ to be a reliably good. The larger this parameter, the smaller the impact of the robust statistics weighting, conversely as this parameter approaches 1, more and more of the image will be affected and potentially causing bright sources to be suppressed.
	
	\begin{figure*}
		\centering
		%\leavevmode
		\includegraphics[width={0.85\linewidth}]{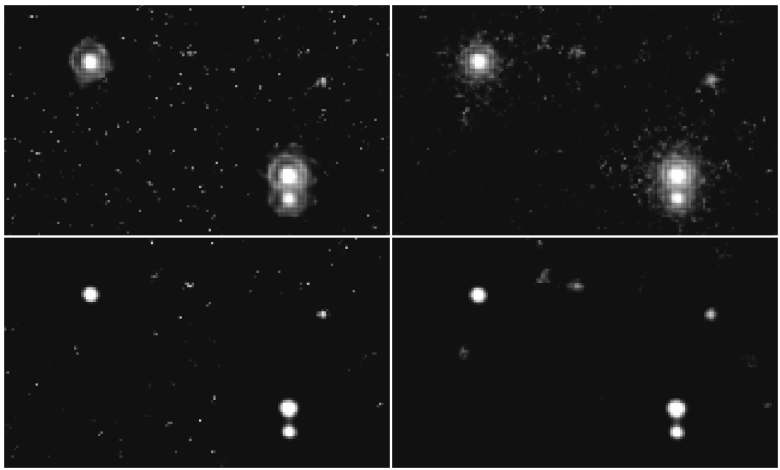}
		\caption{unmodified multiframe blind deconvolution (MFBD) (\emph{top-left}), MFBD with Update Clipping (\emph{top-right}), MFBD with robust statistics weighting (\emph{bottom-left}), MFBD with robust statistics weighting and update clipping (\emph{bottom-right}). The clipping removes much of the background noise, but does not have a large effect on the PSF. Robust Statistics weighting provides a much improved PSF and the reduction of noise around objects.}
		\label{fig:robustCompare}
	\end{figure*}
	
	The sole application of the robust statistics weighting reduces the number of artifacts and noise around and between objects, while also resolving a greatly improved PSF, see \emph{bottom-left} of fig.~\ref{fig:robustCompare}. Notice the lack of halos around the objects, as well as the reduced number of the background speckles. 
	
	To further reduce these background artifacts and prevent erratic updates from disrupting fainter sources, ie. the top center region of the image, we introduce convergence control using the update clipping, section \ref{sec:clipping}. In our testing we found $d=2.0$ to be a good clipping threshold. On its own (\emph{top-right} in fig.~\ref{fig:robustCompare}), the clipping clears up the vast majority of background speckles. 
	\begin{figure*}
		\centering
		%\leavevmode
		\includegraphics[width={0.85\linewidth}]{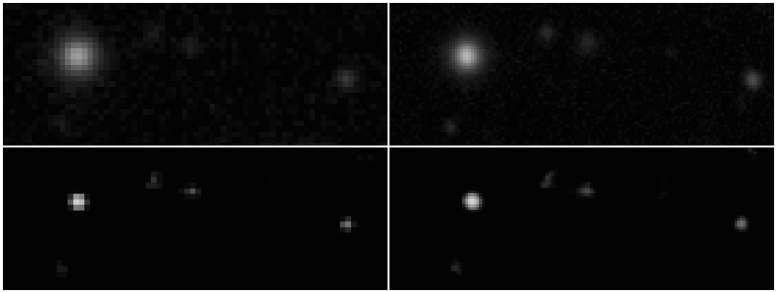}
		\caption{Our standard (\emph{bottom-left}) and super resolution (\emph{bottom-right}) results show a clear improvement in Signal-to-noise as well as a much smaller PSF as compared to both the SDSS Coadd (\emph{top-left}) and CFHTLS Coadd (\emph{top-right}). Our super resolution result produces images in comparable resolution and detail to CFHTLS, which is an impressive achievement given that CFHTLS is a much deeper survey with more than double the resolution.}
		\label{fig:dualStars}
	\end{figure*}
	
	In combination these methods produce clean and sharp images, which far exceed the quality of the SDSS coadd, which was generated using a similar set of input image, and even exceeding the quality of the CFHTLS coadd. For a fair comparison we show an excerpt of our results with and without super resolution enabled to match the SDSS and CFHTLS coadds, see fig.~\ref{fig:dualStars}.
	
\begin{figure}
		\centering
		\includegraphics[width=0.85\linewidth]{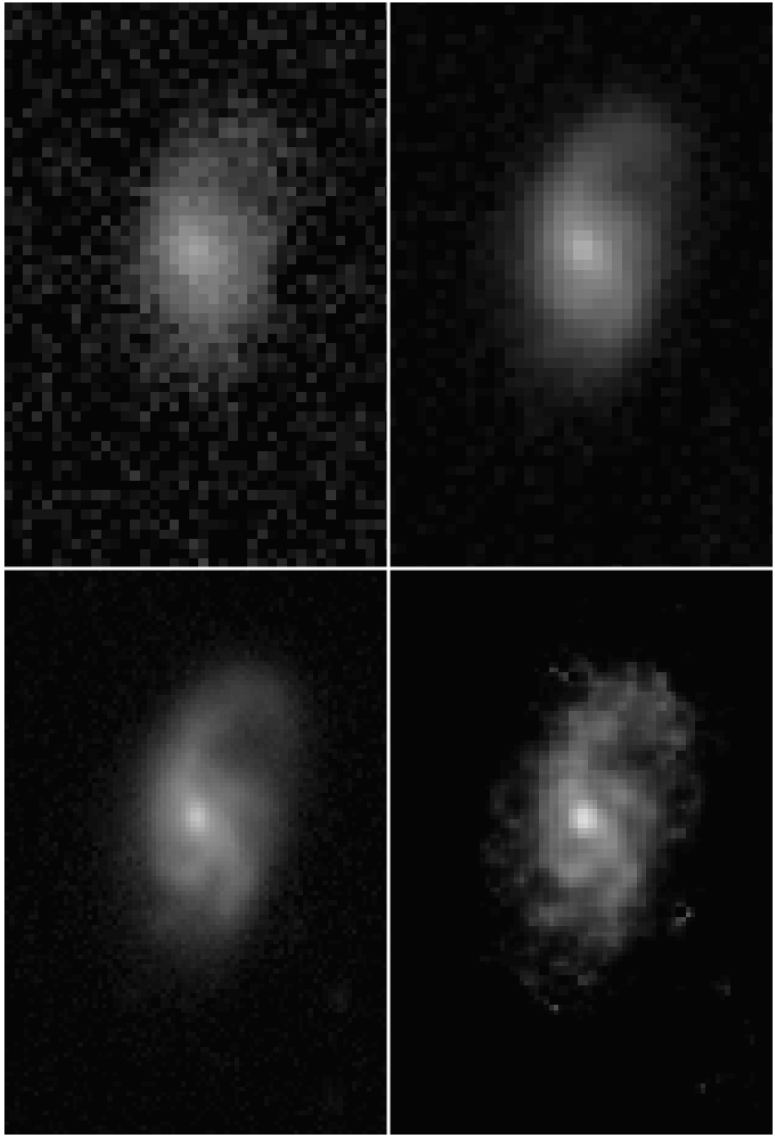}
		\caption{Deconvolution of complex sources such as galaxies are a good benchmark of how well a PSF is formed. Here we compare our result (\emph{bottom-right}) against a typical input frame (\emph{top-left}), as well as the SDSS (\emph{top-right}) and CFHTLS (\emph{bottom-left}) coadds.}
		\label{fig:galaxyCompare}
	\end{figure}

	Point sources are less challenging to deconvolve than complex sources, galaxies are a good benchmark for validating the quality of our PSFs, as there is more structure that would otherwise get washed out by a poor PSF. In fig.~\ref{fig:galaxyCompare} we show our robust performance on a spiral galaxy, note the additional detail available on the spiral arms, also note the similarity in the structures of the CFHT coadd.
	
	\subsection{Software: Implementation and Performance}
	
	An important consideration with approaches like these is their performance. If a great method takes many hours to process a set of small images, it may yield great results, but it is impractical in a real world applications. For that reason part of our focuses has been on implementing GPU-accelerated methods for all of our compute intensive tasks. Originally we planned on implementing both a full CPU and GPU code path for all features, but as we began to investigating larger (2k by 2k) images, especially with super resolution, it became apparent that a CPU implementation would be too slow to be useful.
	
	Our implementation is written in Python, heavily relying on the use of numpy \citep{ascher2001numerical}, a fast numerical and array library, and pyCUDA \citep{kloeckner_pycuda_2012}, a Python interface to NVIDIA's GPU-programming language CUDA \citep{nickolls2008scalable}. Python, while perhaps not quite as performant as C or C++, offers access to an incredibly rich and mature environment of existing scientific libraries, making development and experimentation a pleasure. However, while investigating the addition of wavelet filtering, we found a severe lack of functional GPU accelerated wavelet transform libraries. At the time of writing there seems to be only one GPU-accelerated python wavelet transform library, that being PyGASP \citep{bowman2013pygasp}, which wildly under performs expectations, so much so that the well-known CPU-based python wavelet library, pywt \citep{wasilewski2010pywavelets}, outperformed PyGASP by 5x on the same input. While we ultimately omitted wavelet filtering for this paper, we believe it may be beneficial and is planned as future work. 
	
	Our current implementation can process 140 images (2k by 2k) using standard resolution in under 5 minutes and using 4k by 4k super resolution in approximately 10 minutes, testing performed on an NVIDIA K20 GPU. Of course different parameter settings as well as varying quality of input image does affect the processing time. The times given here are based on our testing completed with SDSS images.

	\section{Future Work and Summary} \label{sec:summary}
	
	While our method performs well in many scenarios, we have identified a variety of issues to be addressed as future work. In order for our method to perform at it's best, we rely on a uniform and known background level. Images with poor background subtraction and/or non-uniform backgrounds can impact the final result by wrongly adding or removing flux. At the moment these effects can be largely mitigated by aggressively clipping our updates. Preferably we would solve for a multi-variate background along side the PSF (as an additional set of parameters), allowing us to appropriately remove the background. This would further improve our results, by letting us dial back the clipping as well as allowing us to include images that previously had to be discarded for large background gradients.
	Another area of exploration is filtering and smoothing of the residual in order to dampen noise. Initially we figured an approach based on Starck and Murtagh's wavelet filtering method \citep{starck1994image} would work well, but after the disappointing results with painfully slow wavelet transform libraries, we decided to revisit this at a later time.
	
	Additionally it would be beneficial to further validate our results against other surveys such as the Dark Energy Survey (DES) \citep{dark2005dark} and the Hubble Legacy Archive (HLA) \citep{jenkner2006concept,whitmore2008source} verify the quality of our results. Also as part of our future work we plan on evaluating our results in terms of changes to astronometric and photometric errors. First results look promising, but more investigation is needed.
	
	Our method shows a vast gain in quality over Coadds, lucky imaging and the standard Multiframe Blind Deconvolution. The robust statistics weighting successfully prevents outliers to overtake the cost function, therefore providing more realistic PSF estimates and much reduced background noise, the update clipping is shown to prevent artifacts in noise dominated areas, as well as preventing erratic updates from breaking up faint sources. In combination with Super Resolution our methods outperform the current state-of-the-art for combining images, allowing us to produce images that exceed even coadds of a much larger telescope with twice the spacial resolution.  We show how to produce robust results from real data in a timely manner.
	Our method will be instrumental in tackling the colossal datasets produced by current and upcoming surveys, enabling the production of higher quality and depth of observations than initially available from the instrument itself.

	\section*{Acknowledgment}

	%The authors are grateful to Rick White for several invaluable discussions about these and related topics. Also 
 	The authors acknowledge valuable initial discussions about blind deconvolution ideas with Michael Hirsch for discussions on related topics.
	This research has been funded by NSF Grant AST-1412566.

	%\ifCLASSOPTIONcaptionsoff
	%\newpage
	%\fi

	\bibliographystyle{asp2010}
	\bibliography{./Streaming_MFBD}

\begin{thebibliography}{}
\expandafter\ifx\csname natexlab\endcsname\relax\def\natexlab#1{#1}\fi
\expandafter\ifx\csname url\endcsname\relax
  \def\url#1{\texttt{#1}}\fi
\expandafter\ifx\csname urlprefix\endcsname\relax\def\urlprefix{URL }\fi
\providecommand{\eprint}[2][]{\url{#2}}

\bibitem[{Abazajian et~al.(2009)Abazajian, Adelman-McCarthy, Ag{\"u}eros,
  Allam, Prieto, An, Anderson, Anderson, Annis, Bahcall
  et~al.}]{abazajian2009seventh}
Abazajian, K.~N., Adelman-McCarthy, J.~K., Ag{\"u}eros, M.~A., Allam, S.~S.,
  Prieto, C.~A., An, D., Anderson, K.~S., Anderson, S.~F., Annis, J., Bahcall,
  N.~A., et~al. 2009, The Astrophysical Journal Supplement Series, 182, 543

\bibitem[{Annis et~al.(2011)Annis, Soares-Santos, Strauss, Becker, Dodelson,
  Fan, Gunn, Hao, Ivezic, Jester et~al.}]{annis2011sdss}
Annis, J., Soares-Santos, M., Strauss, M.~A., Becker, A.~C., Dodelson, S., Fan,
  X., Gunn, J.~E., Hao, J., Ivezic, Z., Jester, S., et~al. 2011, arXiv preprint
  arXiv:1111.6619

\bibitem[{Ascher et~al.(2001)Ascher, Dubois, Hinsen, Hugunin, Oliphant
  et~al.}]{ascher2001numerical}
Ascher, D., Dubois, P.~F., Hinsen, K., Hugunin, J., Oliphant, T., et~al. 2001,
  Numerical python

\bibitem[{Bowman et~al.(2013)Bowman, Carrier, \& Wolffe}]{bowman2013pygasp}
Bowman, N., Carrier, E., \& Wolffe, G. 2013, in Electro/Information Technology
  (EIT), 2013 IEEE International Conference on (IEEE), 1

\bibitem[{Burrows et~al.(1991)Burrows, Holtzman, Faber, Bely, Hasan, Lynds, \&
  Schroeder}]{burrows1991imaging}
Burrows, C.~J., Holtzman, J.~A., Faber, S., Bely, P.~Y., Hasan, H., Lynds, C.,
  \& Schroeder, D. 1991, The Astrophysical Journal, 369, L21

\bibitem[{Byrd et~al.(1995)Byrd, Lu, Nocedal, \& Zhu}]{byrd1995limited}
Byrd, R.~H., Lu, P., Nocedal, J., \& Zhu, C. 1995, SIAM Journal on Scientific
  Computing, 16, 1190

\bibitem[{Collaboration et~al.(2005)}]{dark2005dark}
Collaboration, D. E.~S., et~al. 2005, arXiv preprint astro-ph/0510346

\bibitem[{Cunningham \& Anthony(1993)}]{cunningham1993image}
Cunningham, C.~C., \& Anthony, D. 1993, Icarus, 102, 307

\bibitem[{Fish et~al.(1995)Fish, Brinicombe, Pike, \& Walker}]{fish1995blind}
Fish, D., Brinicombe, A., Pike, E., \& Walker, J. 1995, JOSA A, 12, 58

\bibitem[{Harmeling et~al.(2009)Harmeling, Hirsch, Sra, \&
  Scholkopf}]{harmeling2009online}
Harmeling, S., Hirsch, M., Sra, S., \& Scholkopf, B. 2009, in Computational
  Photography (ICCP), 2009 IEEE International Conference on (IEEE), 1

\bibitem[{Harmeling et~al.(2010)Harmeling, Sra, Hirsch, \&
  Scholkopf}]{harmeling2010multiframe}
Harmeling, S., Sra, S., Hirsch, M., \& Scholkopf, B. 2010, in Image Processing
  (ICIP), 2010 17th IEEE International Conference on (IEEE), 3313

\bibitem[{Homrighausen et~al.(2011)Homrighausen, Genovese, Connolly, Becker, \&
  Owen}]{homrighausen2011image}
Homrighausen, D., Genovese, C., Connolly, A., Becker, A., \& Owen, R. 2011,
  Publications of the Astronomical Society of the Pacific, 123, 1117

\bibitem[{Hudelot et~al.(2012)Hudelot, Goranova, Mellier, McCracken, Magnard,
  Monnerville, Smah, Cuillandre et~al.}]{hudelot2012t0007}
Hudelot, P., Goranova, Y., Mellier, Y., McCracken, H.~J., Magnard, F.,
  Monnerville, M., Smah, G., Cuillandre, J.-C., et~al. 2012, T0007: The final
  cfhtls release

\bibitem[{Jenkner et~al.(2006)Jenkner, Doxsey, Hanisch, Lubow, Miller~III, \&
  White}]{jenkner2006concept}
Jenkner, H., Doxsey, R., Hanisch, R., Lubow, S., Miller~III, W., \& White, R.
  2006, in Astronomical Data Analysis Software and Systems XV, vol. 351, 406

\bibitem[{{Kl{\"o}ckner} et~al.(2012){Kl{\"o}ckner}, {Pinto}, {Lee},
  {Catanzaro}, {Ivanov}, \& {Fasih}}]{kloeckner_pycuda_2012}
{Kl{\"o}ckner}, A., {Pinto}, N., {Lee}, Y., {Catanzaro}, B., {Ivanov}, P., \&
  {Fasih}, A. 2012, Parallel Computing, 38, 157

\bibitem[{Krist \& Hasan(1993)}]{krist1993deconvolution}
Krist, J., \& Hasan, H. 1993, in Astronomical Data Analysis Software and
  Systems II, vol.~52, 530

\bibitem[{Lauer(2002)}]{lauer2002deconvolution}
Lauer, T.~R. 2002, arXiv preprint astro-ph/0208247

\bibitem[{Lucy \& Hook(1992)}]{lucy1992co}
Lucy, L., \& Hook, R. 1992, in Astronomical Data Analysis Software and Systems
  I, vol.~25, 277

\bibitem[{Lucy(1974)}]{lucy1974iterative}
Lucy, L.~B. 1974, The Astronomical Journal, 79, 745

\bibitem[{Maronna et~al.(2006)Maronna, Martin, \& Yohai}]{maronna2006robust}
Maronna, R., Martin, D., \& Yohai, V. 2006, Robust statistics (John Wiley \&
  Sons, Chichester. ISBN)

\bibitem[{Nickolls et~al.(2008)Nickolls, Buck, Garland, \&
  Skadron}]{nickolls2008scalable}
Nickolls, J., Buck, I., Garland, M., \& Skadron, K. 2008, Queue, 6, 40

\bibitem[{Nunez \& Llacer(1993)}]{nunez1993general}
Nunez, J., \& Llacer, J. 1993, Publications of the Astronomical Society of the
  Pacific, 1192

\bibitem[{Richardson(1972)}]{richardson1972bayesian}
Richardson, W.~H. 1972, JOSA, 62, 55

\bibitem[{Starck et~al.(2002)Starck, Pantin, \&
  Murtagh}]{starck2002deconvolution}
Starck, J., Pantin, E., \& Murtagh, F. 2002, Publications of the Astronomical
  Society of the Pacific, 114, 1051

\bibitem[{Starck \& Murtagh(1994)}]{starck1994image}
Starck, J.-L., \& Murtagh, F. 1994, Astronomy and Astrophysics, 288, 342

\bibitem[{Tubbs(2003)}]{tubbs2003lucky}
Tubbs, R.~N. 2003, arXiv preprint astro-ph/0311481

\bibitem[{Wasilewski(2010)}]{wasilewski2010pywavelets}
Wasilewski, F. 2010, Pywavelets: Discrete wavelet transform in python.
  \url{http://www.pybytes.com/pywavelets/}

\bibitem[{White(1994)}]{white1994image}
White, R.~L. 1994, in 1994 Symposium on Astronomical Telescopes \&
  Instrumentation for the 21st Century (International Society for Optics and
  Photonics), 1342

\bibitem[{Whitmore et~al.(2008)Whitmore, Lindsay, \&
  Stankiewicz}]{whitmore2008source}
Whitmore, B., Lindsay, K., \& Stankiewicz, M. 2008, in Astronomical Data
  Analysis Software and Systems XVII, vol. 394, 481

\bibitem[{York et~al.(2000)York, Adelman, Anderson~Jr, Anderson, Annis,
  Bahcall, Bakken, Barkhouser, Bastian, Berman et~al.}]{york2000sloan}
York, D.~G., Adelman, J., Anderson~Jr, J.~E., Anderson, S.~F., Annis, J.,
  Bahcall, N.~A., Bakken, J., Barkhouser, R., Bastian, S., Berman, E., et~al.
  2000, The Astronomical Journal, 120, 1579

\end{thebibliography}
	
\end{document}